\documentclass[showpacs,aps,prb,reprint,superscriptaddress]{revtex4-2}
\usepackage{bm}
\usepackage{graphicx}
\usepackage{amsmath}
\usepackage{amsfonts}
\usepackage{amssymb}
\usepackage{epstopdf}
\usepackage{hyperref}
\usepackage{braket}
\usepackage{float}

\usepackage{enumitem}
\usepackage{url}
\hypersetup{
    colorlinks,%
    citecolor=blue,%
    linkcolor=blue,%
    urlcolor=blue
}
\usepackage{tikz}\usetikzlibrary{petri}

\begin{document}

\newcommand{\be}   {\begin{equation}}
\newcommand{\ee}   {\end{equation}}
\newcommand{\ba}   {\begin{eqnarray}}
\newcommand{\ea}   {\end{eqnarray}}
\newcommand{\ve}  {\varepsilon}

\title{ Electric field-induced edge state oscillations in GaSb/InAs quantum wells. }

\author{Marcos H.~L. de Medeiros}
\affiliation{Instituto de F\'{\i}sica, Universidade de S\~{a}o Paulo,
C.P.\ 66318, 05315--970 S\~{a}o Paulo, SP, Brazil}
\author{Raphael L.~R.~C. Teixeira}
\affiliation{Instituto de F\'{\i}sica, Universidade de S\~{a}o Paulo,
C.P.\ 66318, 05315--970 S\~{a}o Paulo, SP, Brazil}
\author{Guilherme M. Sipahi}
\affiliation{Instituto de F\'{\i}sica de S\~{a}o Carlos, Universidade de S\~{a}o Paulo, 13566-590 S\~{a}o Carlos, S\~{a}o Paulo, Brazil}
\author{Luis G.~G.~V. Dias da Silva}
\affiliation{Instituto de F\'{\i}sica, Universidade de S\~{a}o Paulo,
C.P.\ 66318, 05315--970 S\~{a}o Paulo, SP, Brazil}

\date{ \today }

\begin{abstract}

Inverted-gap GaSb/InAs quantum wells have long been predicted to show quantum spin Hall insulator (QSHI) behavior. The experimental characterization of the QSHI phase in these systems has relied on the presence of quantized edge transport near charge neutrality. However, experimental data showing the presence of edge conductance in the \emph{trivial} regime suggest that additional experimental signatures are needed to characterize the QSHI phase. Here we show that electric field- induced gap oscillations can be used as an indicator of the presence of helical edge states in  system. By studying a realistic low-energy model GaSb/InAs quantum wells derived from $k \cdot p$ band theory, we show that such oscillations are bound to appear in narrow samples as the system is driven to the the the topological phase by the electric field. Our results can serve as a guide for the search of additional experimental signatures of the presence of topologically-protected helical edge states in GaSb/InAs systems.

\end{abstract}

\maketitle

\section{Introduction}
\label{sec:Intro}

Quantum spin Hall insulators (QSHIs) \cite{Kane:Phys.Rev.Lett.:226801:2005,Bernevig:Phys.Rev.Lett.:106802:2006}
  are systems with a bulk band gap and gapless helical edge states, topologically protected from backscattering by the time-reversal symmetry \cite{Wu:106401:2006}. In this sense, QSHI can be thought 
  as two-dimensional versions of 3D topological insulators \cite{hasan2010colloquium,ando2013topological}.  The promises of applications in spintronics and quantum-computing have leveraged the research towards the understanding and the synthesis of such systems.

The main platform for the studies of QSHI behavior are semiconductor heterostructures forming quantum wells of ``inverted-band" materials such as HgTe or GaSb. In fact, QSHI behavior was first predicted to occur in HgTe/CdTe quantum wells \cite{Bernevig.Science.314.1757}, with experimental results consistent with the presence of helical edge states in this system appearing shortly thereafter \cite{Konig:766:2007}. Nevertheless, the fabrication
process of HgTe/CdTe quantum wells presents subtleties making the molecular beam epitaxy growth not broadly accessible. Moreover, mercury compounds are highly toxic and must be handled with extra precautions \cite{Shen:JPCL:22242228:2017}, adding a degree of risk to those involved in the synthesis process.

Later, it was predicted that broken-gap InAs/GaSb asymmetric quantum wells also behave as QSHI \cite{Liu:Phys.Rev.Lett.:236601:2008}. More importantly, 8-band $k \cdot p$ calculations have suggested that the topological transition (i.e., gap inversion)  can be controlled by applying an external electric field along the growth direction \cite{Hu:Phys.Rev.B:94:045317:2016} applied through a potential difference between front- and back-gates. 

These theoretical predictions were then tested in a variety of experiments on InAs/GaSb quantum well samples. Evidence for a gap at charge neutrality \cite{Knez:Phys.Rev.B:201301:2010} and quantized conductance \cite{Knez:Phys.Rev.Lett.:136603:2011, Du:Phys.Rev.Lett.:096802:2015} were reported in small samples (with length $L\lesssim 2 \mu$m). More recently, the topological phase transition as a function of front and back-gate voltages was characterized \cite{Qu:Phys.Rev.Lett.:036803:2015,Nichele:Phys.Rev.Lett.:016801:2017}.

In spite of these experimental developments, finding quantized conductance plateaus in these systems is a challenging task, as the actual conductance values can vary from sample to sample and in different experiments. For instance. the presence of helical sates at zero field in $p$-$n$ junctions at zero magnetic field is inconclusive \cite{Karalic:Phys.Rev.B:241402:2016,Karalic:Phys.Rev.Lett.:206801:2017}.  More importantly, edge state transport has been detected in the \emph{trivial} phase of GaSb/InAs quantum wells  \cite{Nichele:NJP:083005:2016,Nguyen:Phys.Rev.Lett.:077701:2016}. Indeed, such states are seen in several samples and might be contributing to the conductance in the topological phase as well.

This telling example shows that it is not trivial to distinguish non-topological and topologically protected helical states from transport data alone. In a way, this is similar as in the case of Majorana zero modes in semiconductor nanowires  where ``quantized'' values of the conductance can occur even in the absence of topologically-protected modes \cite{DasSarma:Phys.Rev.B:195158:2021}. As such, it is desirable to have additional signatures of the presence of topologically-protected helical edge states in GaSb/InAs samples.

In this work, we address this question by studying a realistic model for GaSb/InAs quantum wells in the presence of an applied electric field and showing that the inter-edge coupling of edge states in narrow samples can lead to oscillations as a function of the the field. Such oscillations occur only in the topological phase and can be directly linked to the presence of helical edge states and thereby, signal the onset of QSHI behavior. As such, the oscillations in the edge states can serve as a confirmation for the presence of topologically-protected helical edge modes in GaSb/InAs quantum well systems.

Moreover, we show that the electric field controls not only the topological transition but also increases the exponential localization of the edge states, similarly to the role played by the  magnetic field in regular quantum Hall edge states. In this sense, the situation is analogous to the energy oscillations seen in other contexts such as Majorana systems \cite{DasSarma:Phys.Rev.B:220506:2012,Chiu:Phys.Rev.B:054504:2017} and zeroth Landau level oscillations in nodal semimetals \cite{Devakul:arXiv2101.05294::2021}.

This paper is organized as follows: in Section \ref{sec:modelmethods} we present ${\bf k}\cdot{\bf p}$ calculations for InAs/GaSb asymmetric quantum wells and then derive a low-energy BHZ-like model with realistic parameters. The resulting band structure, the topological phase transition and the resulting appearance of exponentially localized  edge states in the system are presented in  Section \ref{sec:Results}. More importantly, we show that the edge states' energies and localization length oscillate as function of the applied electric field. Finally, we present our concluding remarks in Section \ref{sec:Conclusions}.

\section{Model}
\label{sec:modelmethods}

\begin{figure}
    \begin{center}
         \includegraphics[width=1\columnwidth]{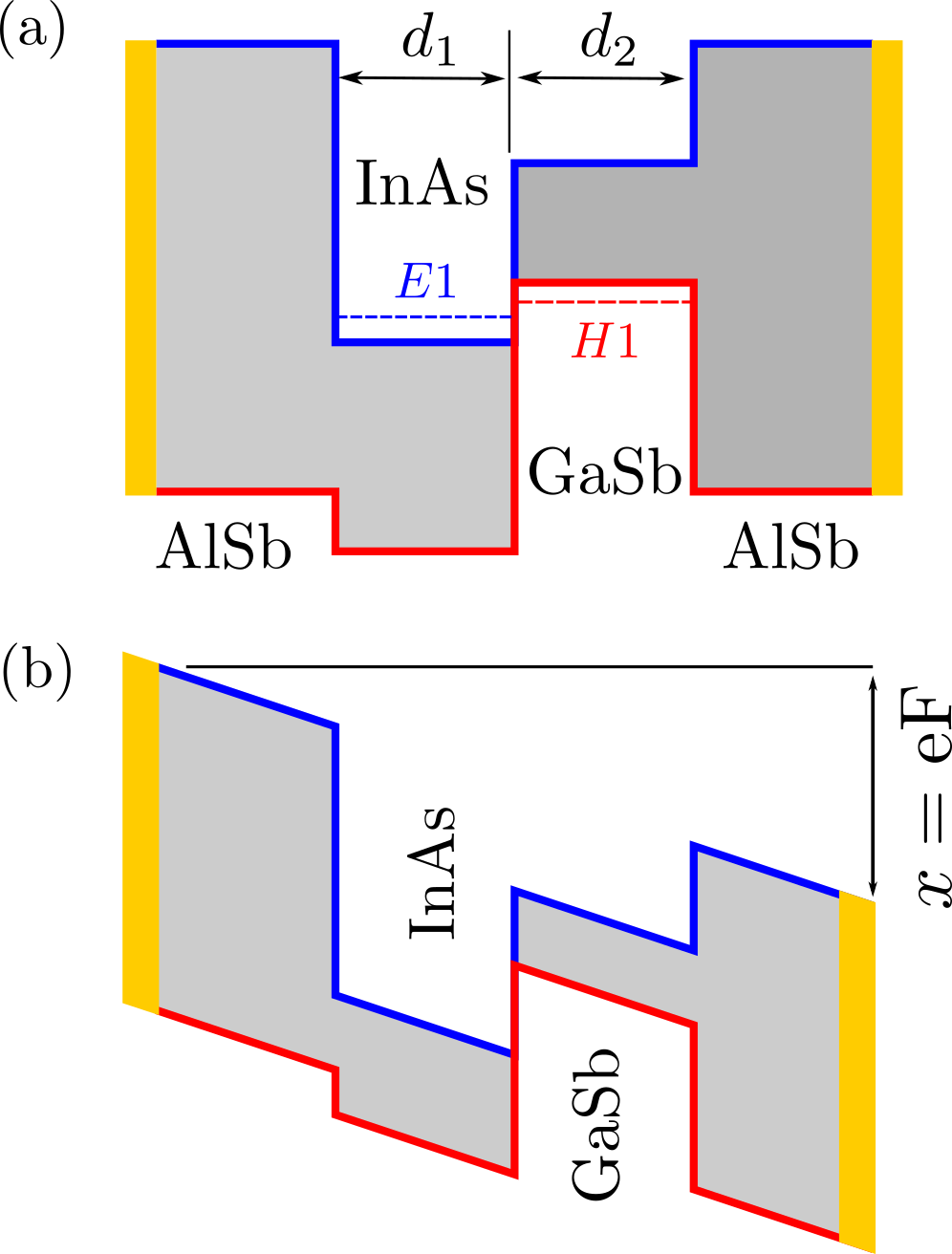}
    \caption{Schematic representation of the asymmetric GaSb/InAs quantum well. 
            (a) Representation of ``inverted'' regime, with the  hole-like state in the GaSb quantum well (width $d_2$) at a higher energy than the electron-like state at the InAs quantum well (width $d_1$). (b)
            Representation of the Stark shift $eF$ referred henceforth simply as ``electric field''. The inverted regime can be obtained also by varying $eF$.
            }
    \label{fig_quantum_well_sketch}
    \end{center}
\end{figure}

We consider the asymmetric InAs/GaSb  system depicted in Figure \ref{fig_quantum_well_sketch}. In this geometry, the system is comprised by a single InAs quantum well in the conduction band (for electron-like states) next to a GaSb valence band quantum well (for hole-like states) leading to a band inversion at the GaSb/InAs interface.
Similarly to the case of HgTe/CdTe \cite{Bernevig.Science.314.1757} and GaSb/InAs  single quantum wells \cite{Liu:Phys.Rev.Lett.:236601:2008,Hu:Phys.Rev.B:94:045317:2016}, a topological phase transition between a trivial insulator and quantum spin Hall phases can be controlled by varying the width $d_1$ of the InAs quantum well \cite{campos2019electrical}. Throughout the paper, we consider fixed widths $d_1$ and $d_2$ corresponding to the quantum spin Hall phase, namely $d_1 = 91$ \AA \ (or 15 monolayers of InAs) and $d_2 = 48.8$ \AA \ (8 monolayers of GaSb). As an additional consistency check, the calculations were performed for  $d_1 = 97$ \AA \ (16 monolayers of InAs), yielding similar results (see Table \ref{table:nonlin}).

The low-energy BHZ Hamiltonian used in this work was obtained following a four-step process. which we now summarize. First, the GaSb/InSb/AlSb system is modeled with an 8-band Kane Hamiltonian properly parametrized. Next, a low-energy effective Hamiltonian is determined with a ``folding-down" procedure and, from both the low-energy and the original 8-band Hamiltonians, the effect of the applied electric field in the system is introduced. Finally, the low-energy Hamiltonian is reviewed and parametrized in order to account the effects of the applied electric field. We discuss these steps in detail in the following sections. 


\subsection{${\bf k}\cdot{\bf p}$ Hamiltonian}
\label{sec:kphamiltonian}

We start from a well-known eight-band Kane Hamiltonian~\cite{kane1966k}, parametrized for the the InAs, GaSb and AlSb bulk alloys. For the modeling of the heterostructure, the confinement of the quantum well in the growth ($z$) direction is included by considering the envelope function approximation (EFA), where the Kane model parameters are taken as $z$-dependent and the substitution $k_z \rightarrow -i\partial_z$ is made. We used a reciprocal space approach where the envelope function is solved by expanding the growth direction into the Fourier coefficients of the potential and $z$-dependent parameters~\cite{Rodrigues:APL:76.105,Rodrigues:JAP:101.113706}. To describe each layer, we used realistic Kane model parameters depicted in Ref. ~\cite{Bastos:JAP:123.065702}.

Due to the confinement in the z-direction, one can assume that the band structure along that direction is flat and that the eigenfunction of the states at $\Gamma$ are a good description of all the functions along the $\Gamma$-Z direction.

\begin{figure}
	\begin{center}
		\includegraphics[width=1\columnwidth]{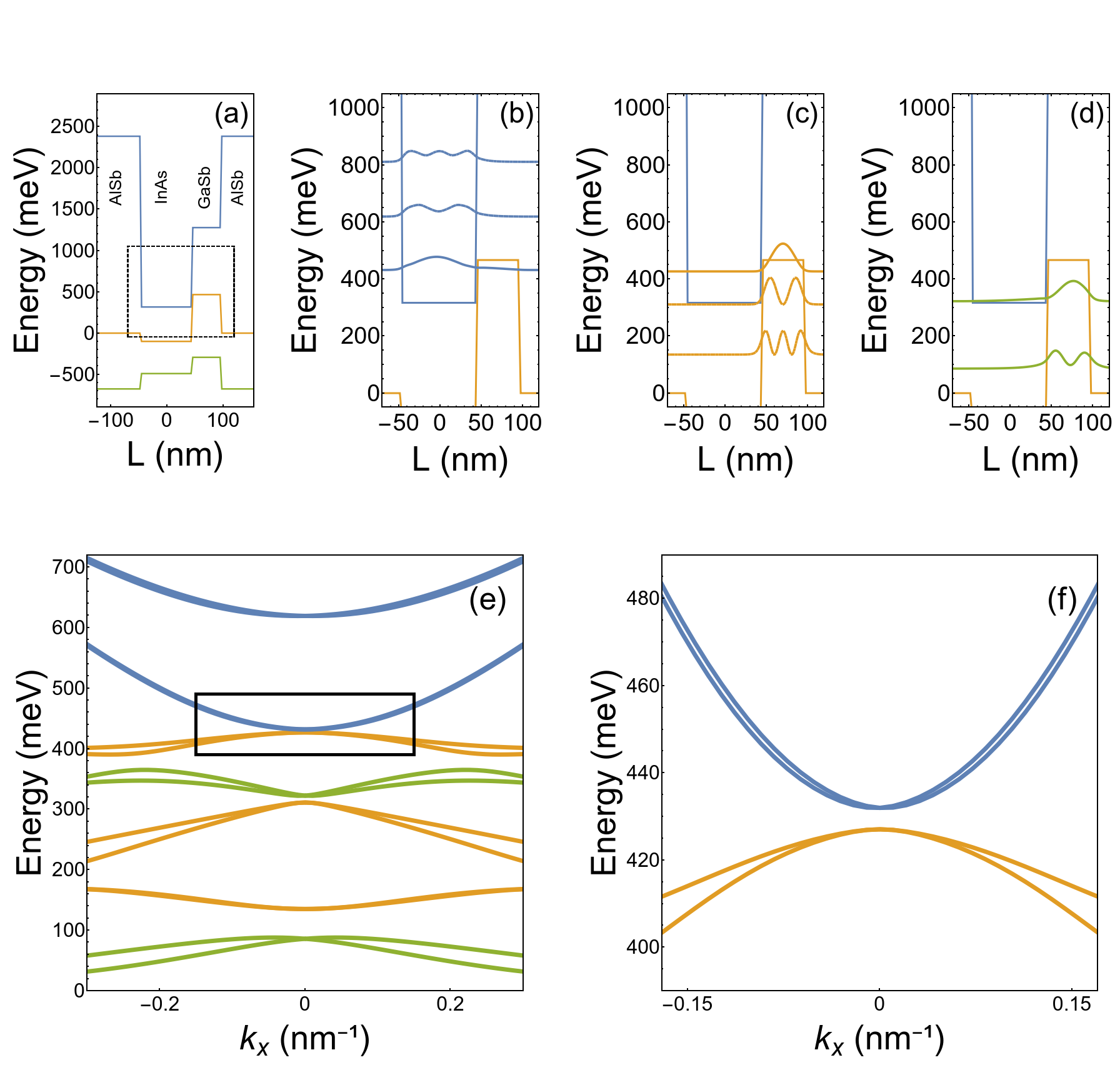}
		\caption{
			Results from the ${\bf k}\cdot{\bf p}$ calculations. In (a) we show the schematic view of the layered system used in the $8 \times 8$ k.p calculation. The dashed box in panel shows the energy range of the states around the Fermi energy that is used in panels (b), (c) and (d) that show the less energetic conduction band, heavy and light hole states, respectively. Panel e) presents in the same energy range the the full $8 \times 8$ calculation along the parallel direction. The box in this panel presents the target band structure used to obtain the four-band effective model, shown in (f). Heavy hole, light hole and electron states are presented in blue, yellow and green respectively. 
		}
		\label{FromKane2BHZ}
	\end{center}
\end{figure}


Fig.~\ref{FromKane2BHZ} presents the system band profiles, wave functions  and  band structures. Panel (a) shows the band profiles for conduction and valence band. The box indicates the energy range around Fermi level inspected in this analysis. Panels (b), (c) and (d) present the density probabilities of the first conduction band, heavy and light hole states, respectively. 
The band structure associated with these  lowest states in the [100]-direction is depicted in panel (e), and panel (f) details the 4 states that may be inverted under the application of the electric field along the growth direction in the region defined by the box in panel (e). The goal of our modeling is to define a realistic simplified Hamiltonian that reproduces the band structure presented on the last panel.

\subsection{Projected perturbation method}
\label{sec:projectedMethod}

After solving for the eigenenergies at the $\Gamma$ point in momentum space ($k_x = k_y = 0$), 
\begin{equation}
    \mathcal{H}^0\Ket{\varphi_{n}} = E_n\Ket{\varphi_n} \; ,
\end{equation}
we have selected the eigenstates $\ket{\varphi}$ most affected by the inversion of bands, i.e., where the mixing of conduction bands and valence bands would be more important. The influence of the states that does not belong to this set can be shown to be small in first order by the L\"owdin \cite{lowdin1951note} perturbation theory.

Therefore the full analytical $8 \times8$ Kane Hamiltonian expanded into plane waves in the $z$-direction is projected over those selected states at the $\Gamma$-point, resulting in an effective 2D-Hamiltonian
\begin{equation}
    H = (\Bra{\varphi_1},\cdots,\Bra{\varphi_{16}})\mathcal{H}(\mathbf{k})
    \begin{pmatrix}
        \Ket{\varphi_1}\\
        \vdots\\
        \Ket{\varphi_{16}}
    \end{pmatrix},
\end{equation}
where we have taken eight doubly degenerate bands due to spin degeneracy presented in Figs.~\ref{FromKane2BHZ}(b), (c) and (d).

The chosen 16 states are the ones most affected by the band inversion, where the mixing of CB and VB is important. Using the L\"owdin perturbation scheme~\cite{lowdin1951note} to validate this set, we chose the set of states mostly affected by the inversion as our unperturbed set in which the first order corrections are small. 
Finally, the band structure of the 16-state Hamiltonian seemingly compares to the one found with the much more expensive envelope function ${8\,\text{NPW}\times8\,\text{NPW}}$ Kane Hamiltonian, where NPW is the number of plane waves of the expansion.

\subsection{Effective low-energy Hamiltonian}
\label{sec:effectivemodel}

Coming back to the L\"owdin perturbation scheme, since we have a Hamiltonian that fully describes our problem, we can still reduce it by defining a new set of unperturbed functions. We chose them as the functions in the small box in Fig.~\ref*{FromKane2BHZ}~e) and apply the first order correction using the other twelve states in a numeric folding down procedure by using the Schur's complement \cite{Zhang_SchurComplement_2005}. 
The final result is a $4\times4$ Hamiltonian matrix with the format
\begin{equation}
    \label{eq_H_97_field_dependent}
    H =
    \begin{bmatrix}
        H_{+}(\mathbf{k})   &  H_{\pm}(\mathbf{k}) \\
        H_{\mp}(\mathbf{k})  &  H_{-}(\mathbf{k})
    \end{bmatrix},
\end{equation}
where the matrix $H_{+}$ is defined as 
\begin{equation}
    \label{eq_H_97_plus} 
    H_{+}(\mathbf{k}) =
    \begin{bmatrix}
     \varepsilon_c (\mathbf{k})  &   iP\mathbf{k}_{+}  \\
    -iP\mathbf{k}_{-}	&  \varepsilon_{v} (\mathbf{k}) \\
     \end{bmatrix},
\end{equation}
with
\begin{eqnarray}
\varepsilon_c(\mathbf{k})= E_c +\alpha_c \mathbf{k} + \gamma_{c} \mathbf{k}^2 \label{eq:ec}\\
\varepsilon_v(\mathbf{k})= E_v +\alpha_v \mathbf{k} + \gamma_{v} \mathbf{k}^2 \label{eq:ev}
\end{eqnarray}
and the time-reversal symmetry guaranteeing that $H_{-}(\mathbf{k}) = H_{+}^{*}(-\mathbf{k})$. The coupling matrices $H_{\pm}(\mathbf{k})$ and
$H_{\mp}(\mathbf{k})$ are given by
\begin{equation}
    \label{eq_H_97_plus_minus}
    H_{\pm}(\mathbf{k}) =
    \begin{bmatrix}
        0   &   & N_{-}(\mathbf{k}) \\
        N_{+}^{*}(\mathbf{k})   &   0  \\
    \end{bmatrix},
\end{equation}
noticing that $H_{\mp} = (H_{\pm})^{\dagger}$ due to the unitarity. The non-zero elements of coupling matrices are given by
\begin{equation}
    \label{eq_coupling_elements_2x2}
    N_{\pm}(\mathbf{k}) = -(k_x^2 - k_y^2)\eta_2 \pm k_xk_y\eta_3.
\end{equation}

A closer inspection of the  $\eta_{2}$ and $\eta_{3}$ values shows that the folded down off-diagonal terms in the $N_\pm$ blocks are responsible for corrections of the order of 0.01 meV in the region of the fitting ($\left|k_x\right| < 0.15 \textrm{\ nm}^{-1}$), 
having little or no impact at the band structures when compared to the case where we consider the bottom of conduction and top of the valence band states in the full Kane calculation. 
As such, the diagonal approximation ($\eta_2\!=\!\eta_3\!=\!0$) turns out to give an excellent description of the low-energy physics around the $\Gamma$ point.

\begin{figure}
	\begin{center}
		\includegraphics[width=1\columnwidth]{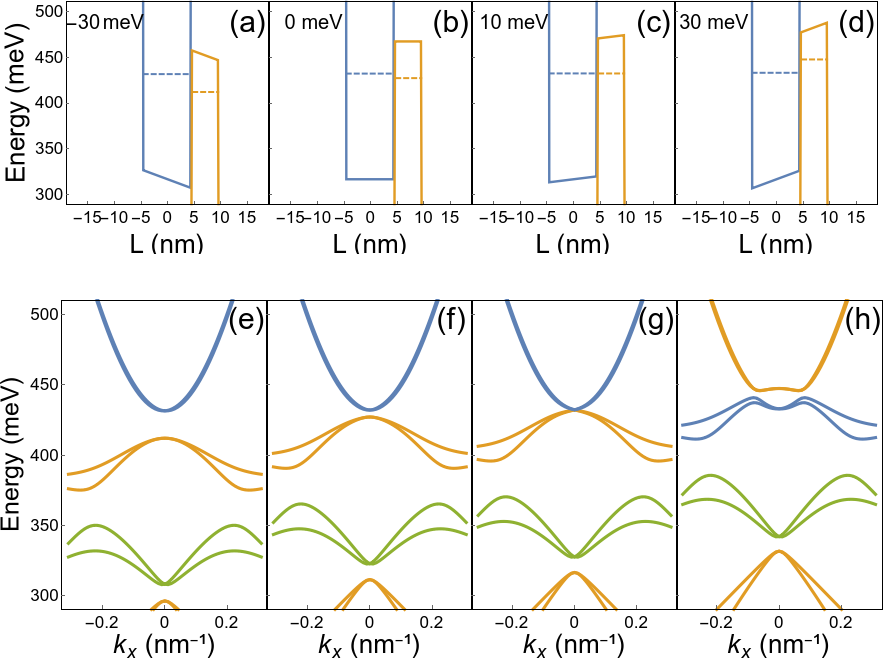}
			\caption{
			Panels (a), (b), (c) and (d) show the Schematic view of the confinement profiles including the heterostructure profile and the applied electric field creating a Stark shift between the interfaces with the AlSb layers on both sides of -30, 0, 10 and 30 meV, respectively. The bottom of the conduction band (blue, on the left) and the top of the valence band (orange, on the right) as well as the highest valence band and lowest conduction band states are shown. (e), (f), (g) and (h) panels show the band structures with the different applied fields. 
		}
		\label{fig_quantum_well_bandstructure}
	\end{center}
\end{figure}


\subsection{Four-band BHZ model}
\label{sec:BHZmodel}

Following these results, we opted for using a BHZ-like Hamiltonian~\cite{Bernevig.Science.314.1757} given by
\begin{equation}
\label{eq:HBHZ}
H_{\rm BHZ} =
\begin{bmatrix}
\hat{H}_{2 \times 2}(\mathbf{k})   &  0 \\
0  &  \hat{H}^{*}_{2 \times 2}(-\mathbf{k})
\end{bmatrix} \; ,
\end{equation}
with
\begin{equation}
\label{eq_2x2_hamiltonian}
    \hat{H}_{2\times2}(\mathbf{k}) =\begin{bmatrix}
    \varepsilon_c(\mathbf{k}) & iP\mathbf{k}_{+} \\
    -iP\mathbf{k}_{-} & \varepsilon_v{\mathbf{k}}
    \end{bmatrix} \; ,
\end{equation}
with $\varepsilon_c(\mathbf{k})$ and $\varepsilon_v(\mathbf{k})$ defined in Eqs.\ \eqref{eq:ec} and \eqref{eq:ev}. The basis set is defined in the usual order \cite{Bernevig.Science.314.1757} as
\begin{equation}
\ket{E,+},~\ket{H,+},~\ket{E,-},~\ket{H,-} \;,
\end{equation}
defined by the character of the states of the Kane model at $\Gamma$-point heavy holes for $\ket{H,\pm}$ and a composition of conduction band electrons (mostly),  light (smaller) and split-off holes (negligible) for $\ket{E,\pm}$.

\subsection{Applied Electric field}
\label{sec:electricfield}

The effect of applied electric field's potential drop across the $z$-directions is shown in Figure~\ref{fig_quantum_well_bandstructure}, both in the potential profile and in the Kane model band structures. A positive drop causes the inversion of conduction and heavy hole bands and a negative one increases the gap. 

Figs.~\ref{fig_quantum_well_bandstructure} (a)--(d) show the potential profiles, together with the $\Gamma$-point energies, across the topological phase transition. As the energy difference between HH and EL states become smaller [Figs.~\ref{fig_quantum_well_bandstructure}  (a),(b)], the gap closes [Figs.~\ref{fig_quantum_well_bandstructure} (c)] and reopens [Figs.~\ref{fig_quantum_well_bandstructure} (d)] with an inverted gap. The respective band structures show usual semiconductor behavior  [Figs.~\ref{fig_quantum_well_bandstructure}(e) and (f)], a gap closure  [Fig.~\ref{fig_quantum_well_bandstructure}(g)] and a ``gapped semimetal"  [Fig.~\ref{fig_quantum_well_bandstructure}(h)].  

\begin{figure}
    \begin{center}
         \includegraphics[width=1\columnwidth]{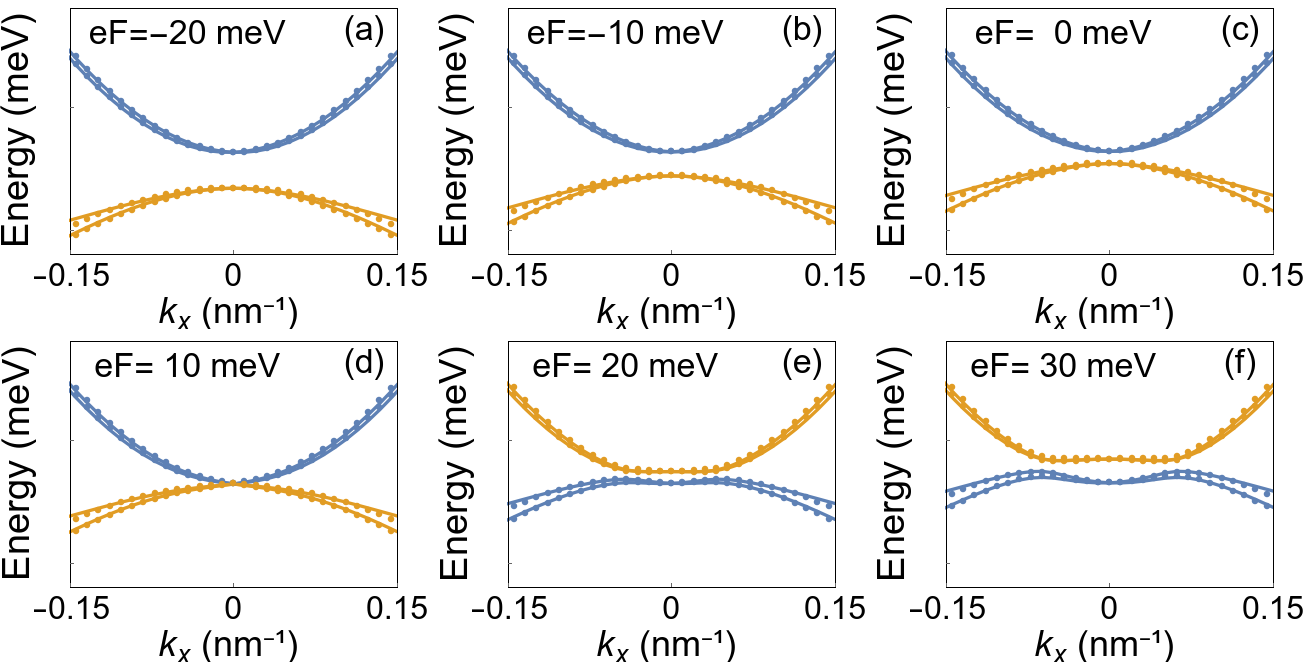}
   \caption{Band structures of the BHZ (dotted lines) and $8\times 8$ k.p (solid lines) models under the influence of eF. From a) to f), eF = -20, -10, 0, 10, 20 and 30 meV, respectively. Yellow (blue) bands have hole-like (electron-like) character at $\Gamma$-point. }
    \label{fig_quantum_well_LowEnergyBands}
    \end{center}
\end{figure}

In the next step we proceeded to the fitting of the BHZ model, with different electric field profiles, in the range from $-30$ to $70$ meV. The fittings of selected systems are presented on Figure~\ref{fig_quantum_well_LowEnergyBands}. From the fittings, one may conclude that the BHZ Hamiltonian provides all the features necessary to the analysis of these four states under the application of an electric field.

\begin{figure}[t]
	\begin{center}
		\includegraphics[width=1\columnwidth]{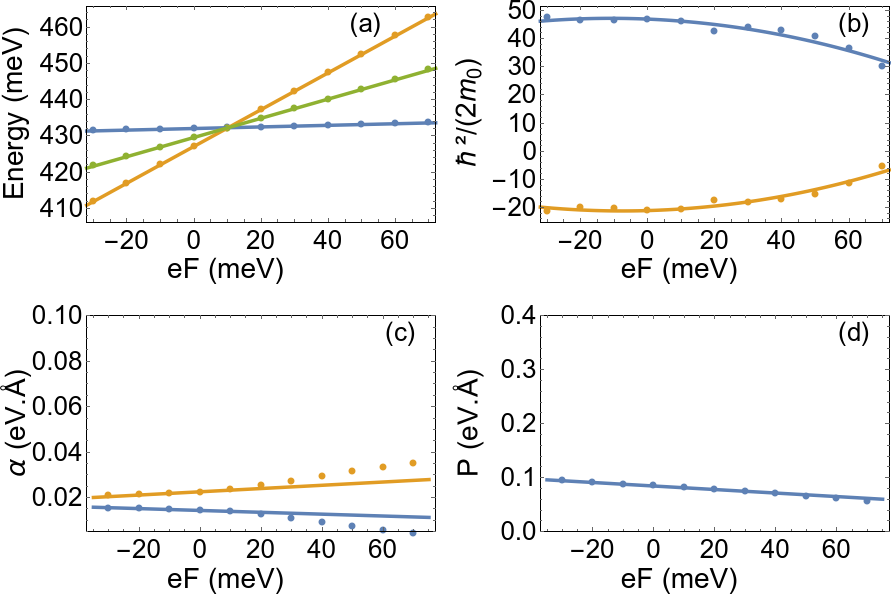}
		\caption{
			Fitting of the variation of BHZ relevant parameters under influence of eF. In solid lines the fitted curves and in dots, the original $8 \times 8$ Hamiltonian fitted values. a) valence band maximum (orange), conduction band minimum (blue) and Fermi level (green); b) effective mass parameter for valence (orange) and conduction (blue) bands; c) $\alpha$ parameter for valence (orange) and conduction (blue) bands; and d) interband interaction parameter $P$.
			}
		\label{fig_quantum_well_ParameterFitting2}
	\end{center}
\end{figure}

As a final step to parameterize our system, we proceed to the fitting of the curves of the relevant parameters under the influence of the applied electric field.  The  dependence of the parameters in Eq.~\eqref{eq:HBHZ} on the electric field $eF$ is given by the following expressions:
\begin{eqnarray}
    E_{(c,v)} (eF)  &=& A_{(c,v)} .eF + B_{(c,v)} \nonumber\\
    \gamma_{(c,v)} (eF)  &=& C_{(c,v)} .eF + D_{(c,v)} \nonumber\\
    \alpha_{(c,v)} (eF)  &=& F_{(c,v)} .eF + G_{(c,v)} \nonumber\\
    P (eF)  &=& p_0 + p_1 .eF \nonumber\\
    \label{eq_Hamiltonian_parametric_functions}
\end{eqnarray}
with the fittings are presented in fig.~\ref{fig_quantum_well_ParameterFitting2}.
Each of these parameters depend on the quantum well thickness, as show in Table \ref{table:nonlin}. Notice that the linear coefficients $\alpha_{c,v}$ are essentially two orders of magnitude smaller than the other relevant terms in the range  $0 \leq eF \lesssim 60$ meV. As such, the linear terms of type $\alpha_{c,v} \mathbf{k}$ can be safely neglected near the $\Gamma$-point for all field values considered.

%
 \begingroup
     \setlength{\tabcolsep}{9pt} 
     \renewcommand{\arraystretch}{1.5} 
     \begin{table}
         \centering 
         \begin{tabular}{c c c c} 
             \hline\hline 
              & $d_1=91$ \AA     & $d_1=97$ \AA      \\ [0.5ex] 
             \hline 
             $A_c$     & $1.613 \times 10^{-6}$   & $9.000 \times 10^{-5}$      \\
             $A_v$     & $3.729 \times 10^{-5}$   & $3.644 \times 10^{-5}$      \\
             $B_c$    & $3.177 \times 10^{-2}$   & $3.109 \times 10^{-2}$      \\
             $B_v$    & $3.140 \times 10^{-2}$   & $3.136 \times 10^{-2}$      \\
             $C_c$     & $-5.007 \times 10^{-2}$  & $-5.379 \times 10^{-2}$      \\
             $C_v$     & $3.480 \times 10^{-2}$   & $5.680 \times 10^{-2}$      \\
             $D_c$     & $4.691 \times 10^{1}$    & $4.625 \times 10^{1}$      \\
             $D_v$     & $-2.105 \times 10^{1}$   & $-1.778 \times 10^{1}$      \\
             $F_c$     & $1.952 \times 10^{-1}$   & $1.318 \times 10^{-1}$      \\
             $F_v$     & $3.068 \times 10^{-1}$   & $4.260 \times 10^{-1}$      \\
             $G_c$     & $-5.545 \times 10^{-4}$  & $-5.419 \times 10^{-4}$      \\
             $G_v$     & $9.703 \times 10^{-4}$   & $8.421 \times 10^{-4}$      \\
             $p_0$    & $8.436 \times 10^{-2}$    & $7.429 \times 10^{-2}$      \\
             $p_1$     & $-3.727 \times 10^{-4}$  & $-3.863 \times 10^{-4}$      \\
             \hline 
         \end{tabular}
         \caption{BHZ Hamiltonian parameters extracted from the fittings of Figs. \ref{fig_quantum_well_LowEnergyBands} and \ref{fig_quantum_well_ParameterFitting2}. Energies are expressed in eV,  lengths in \AA and electric field in V.} 
         \label{table:nonlin} 
     \end{table}
 \endgroup

\section{Topological edge states}
\label{sec:Results}

Once the parametrization of the low-energy BHZ Hamiltonian has been established, we turn to the topological transition and the edge states. To this end, we work with real-space discretization of the $\hat{H}_{2\times 2}$ block written as:
\begin{equation}
    \hat{H}^{\prime}_{2\times2} =\begin{bmatrix}
    \varepsilon^{\prime}_c(\mathbf{k}) & iP\mathbf{k}_{+} \\
    -iP\mathbf{k}_{-} & \varepsilon^{\prime}_v{\mathbf{k}}
    \end{bmatrix}
    \label{eq_Hamiltonian_kwant}
\end{equation}
where the primed diagonal elements are given in terms of the parametric functions defined in Eq. \eqref{eq_Hamiltonian_parametric_functions} (and $\alpha_c \!=\! \alpha_v\! = \! 0$ as previously justified) by
\begin{equation}
    \varepsilon^{\prime}_{(c,v)}(\mathbf{k}) = E_{(c,v)} + \gamma_{(c,v)}\mathbf{k}^2 \; ,
    \label{eq_diagonal_elements_wo_alpha}
\end{equation}
with the parameters set for the case of $d_1=91$ \AA (InAs quantum well width) shown in Table \ref{table:nonlin}.

In the following, we consider infinite strips with translational symmetry in $x$-direction and hard-wall boundary conditions in $y$-direction with width $L_y$. For concreteness, we focus on narrow ($L_y\!=\! 100$ nm) and wide ($L_y\!=\! 200$ nm) systems The calculations are performed with the Kwant package \cite{groth2014kwant}.

\begin{figure}[t]
    \begin{center}
        \includegraphics[width=1.0\columnwidth]{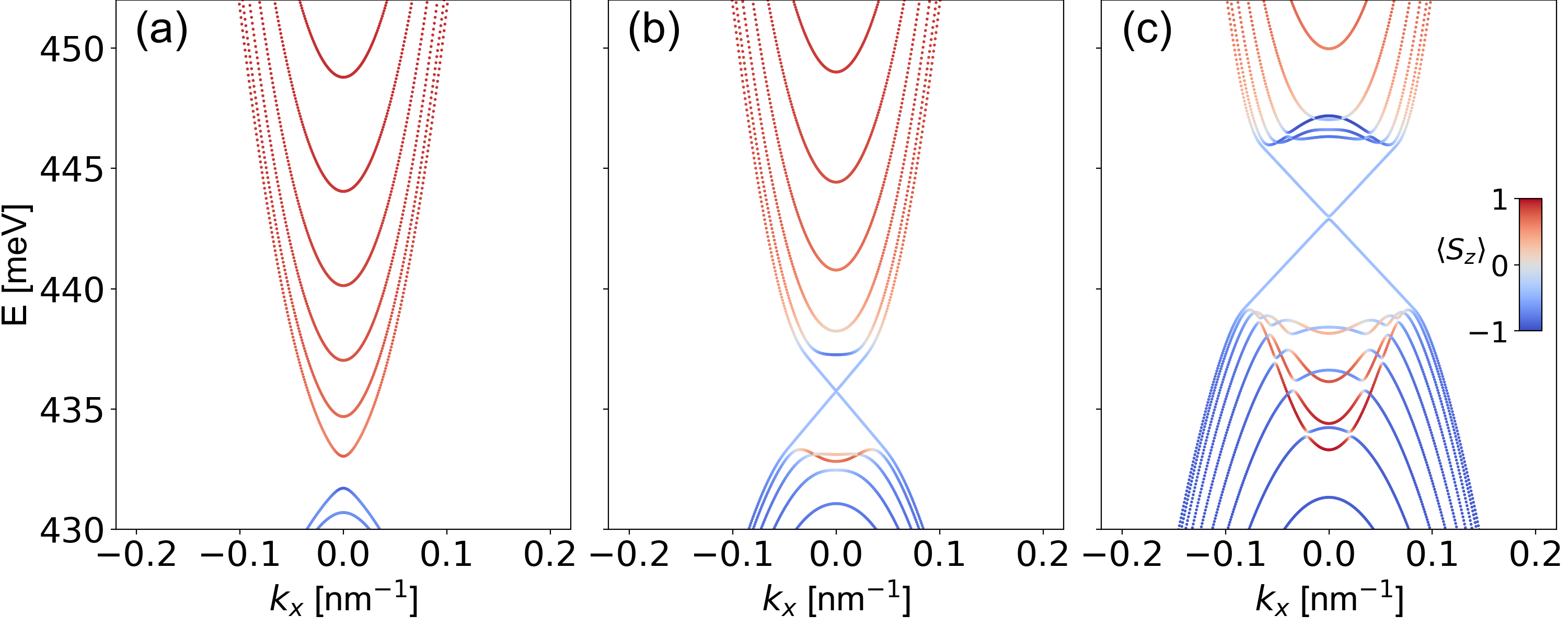}
        \caption{Electric field-driven topological phase transition for the GaSb/InAs double quantum well with $d_1 = 91$ \AA~ and $d_2 = 48.8$ \AA. BHZ energy spectra for $eF=$ (a) 10 meV, (b) 20 meV, and (c) 40 meV. The color bar corresponds to the pseudospin projection. Band inversion and topologically protected edge states are shown in panels (b) and (c).} 
        \label{fig:band_structure_vsk}
    \end{center}
\end{figure}

\subsection{Electric-field driven topological transition}

We begin by characterizing the topological phase transition as a function of the Stark shift energy $eF$  \cite{Hu:Phys.Rev.B:94:045317:2016}.  Figure \ref{fig:band_structure_vsk} shows results for spectrum of the discretized Hamiltonian for different values of $eF$  for wide strips ($L_y\!=\!200$ nm).

The topological transition at $eF \approx 12$ meV is marked by the closing of the gap and subsequent band inversion, along with the the appearance of edge states with linear dispersion (seen in Fig. \ref{fig:band_structure_vsk}-b and \ref{fig:band_structure_vsk}-c). The band inversion can be quantified by the $z$-component of the ``pseudospin'', defined by $\langle S_z \rangle = \int (|\psi_e|^2 - |\psi_h|^2$). As such, the color of each state indicates its composition of the states in terms of $\ket{E,+}$ and $\ket{H,+}$ components.

\begin{figure}[t]
    \begin{center}
        \includegraphics[width=1.0\columnwidth]{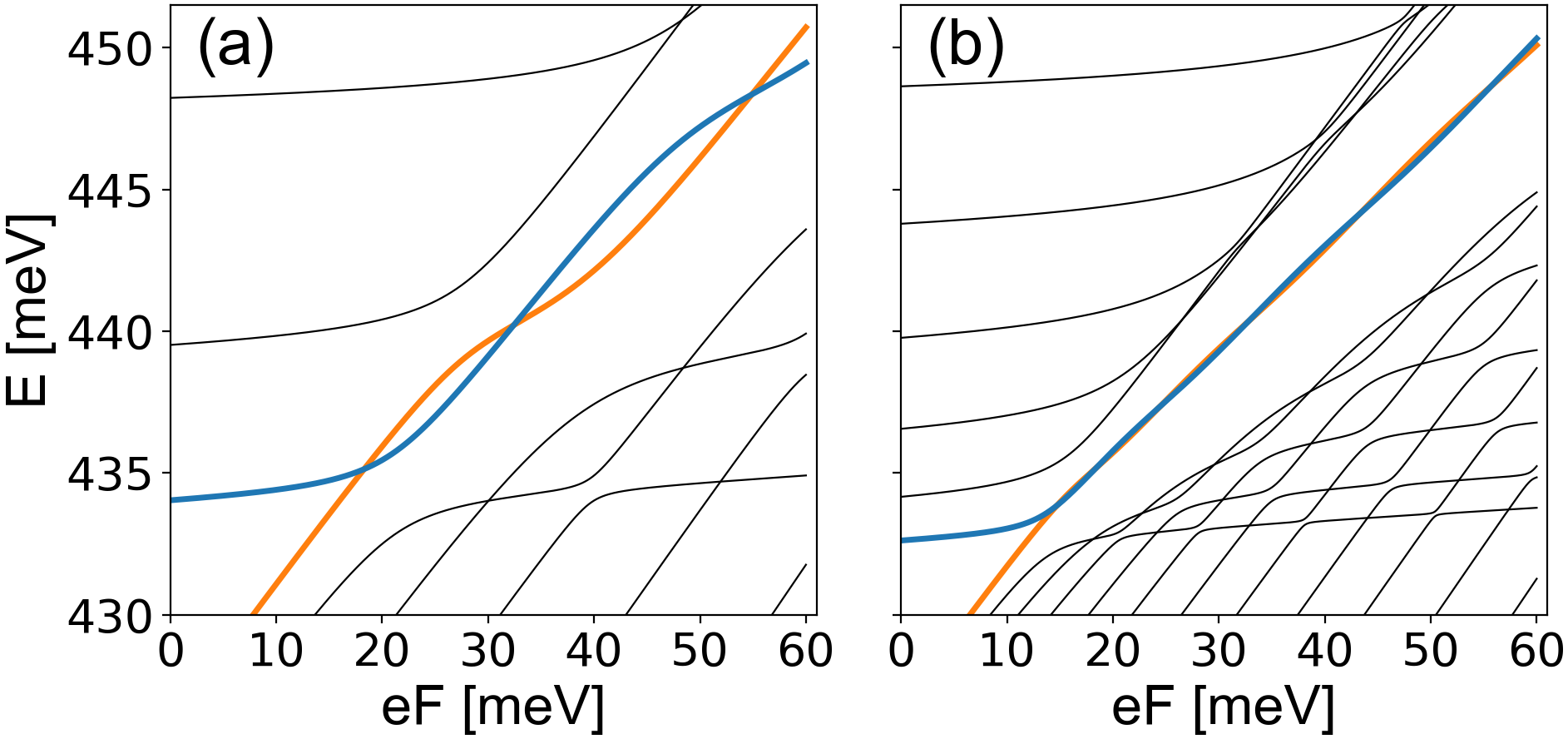}
        \caption{Energy spectra for at $k_x = 0$ versus the electric field for quantum well widths of $L_y = $ (a) 100 nm. and (b) 200 nm. Lines in blue and orange represent the low-lying states, which become the edge states after the transition. Note the oscillations arising from inter-edge coupling for narrow quantum wells.}
        \label{fig:band_structure_keq0}
    \end{center}
\end{figure}

\subsection{Edge state energy oscillations}

Next, we turn to the behavior of the edge states in the topological phase as the Stark shift is increased. Figure \ref{fig:band_structure_keq0} shows the spectrum at $k_x\!=\!0$ as a function of $eF$ for two different values of $L_y$: (a) 100 nm (narrow strip) and (b) 200 nm (wide strip).

The electric-field induced topological phase transition is clearly seen in both cases, marked by a crossing of the states at the top of the the conduction band and at the top of the valence band. These \emph{low-lying energy states} become the sub-gap edge states in the topological phase. 

More importantly,  Fig. \ref{fig:band_structure_keq0} shows a clear oscillatory pattern in the energy of the edge states as a function of $eF$. These oscillations are more pronounced in the case of narrow strips (Fig. \ref{fig:band_structure_keq0}-a). 

The origin of such electric-field-driven oscillations, as it will become clear later, is the inter-edge coupling of the edge states localized at opposite edges. In a sense, these are equivalent to the magnetic field-driven oscillations appearing in short topological nanowires due to the coupling of Majorana zero modes at its ends \cite{DasSarma:Phys.Rev.B:220506:2012,Chiu:Phys.Rev.B:054504:2017}.  As such, these low-lying oscillations occur only in the topological phase and can be regarded as true signatures of the presence of topological edge modes.

\subsection{Edge state localization}

The formal connection between the electric-field-driven edge state oscillations discussed above and those appearing in Majorana systems as a function of magnetic field can further explored in order to get a better understanding the behavior of the edge states as the electric field increases. As such, inspired by Ref.~\cite{DasSarma:Phys.Rev.B:220506:2012}, we propose an ansatz of an oscillatory function with an exponential decay for the edge state wave-functions
\begin{equation}
    |\psi_{b(t)}(y)|^2 \propto e^{-\frac{2\tilde{y}_{b(t)}}{\xi}} \sin^2[k_f \tilde{y}_{b(t)}].
    \label{eq:absolute_squared_wave_function_model}
\end{equation}
where the localization length $\xi$ and the wave number $k_f$ both depend on the electric field (see full expressions in Appendix \ref{sec:appendix_analytical_solution}).

\begin{figure}[t]
    \begin{center}
        \includegraphics[width=1.0\columnwidth]{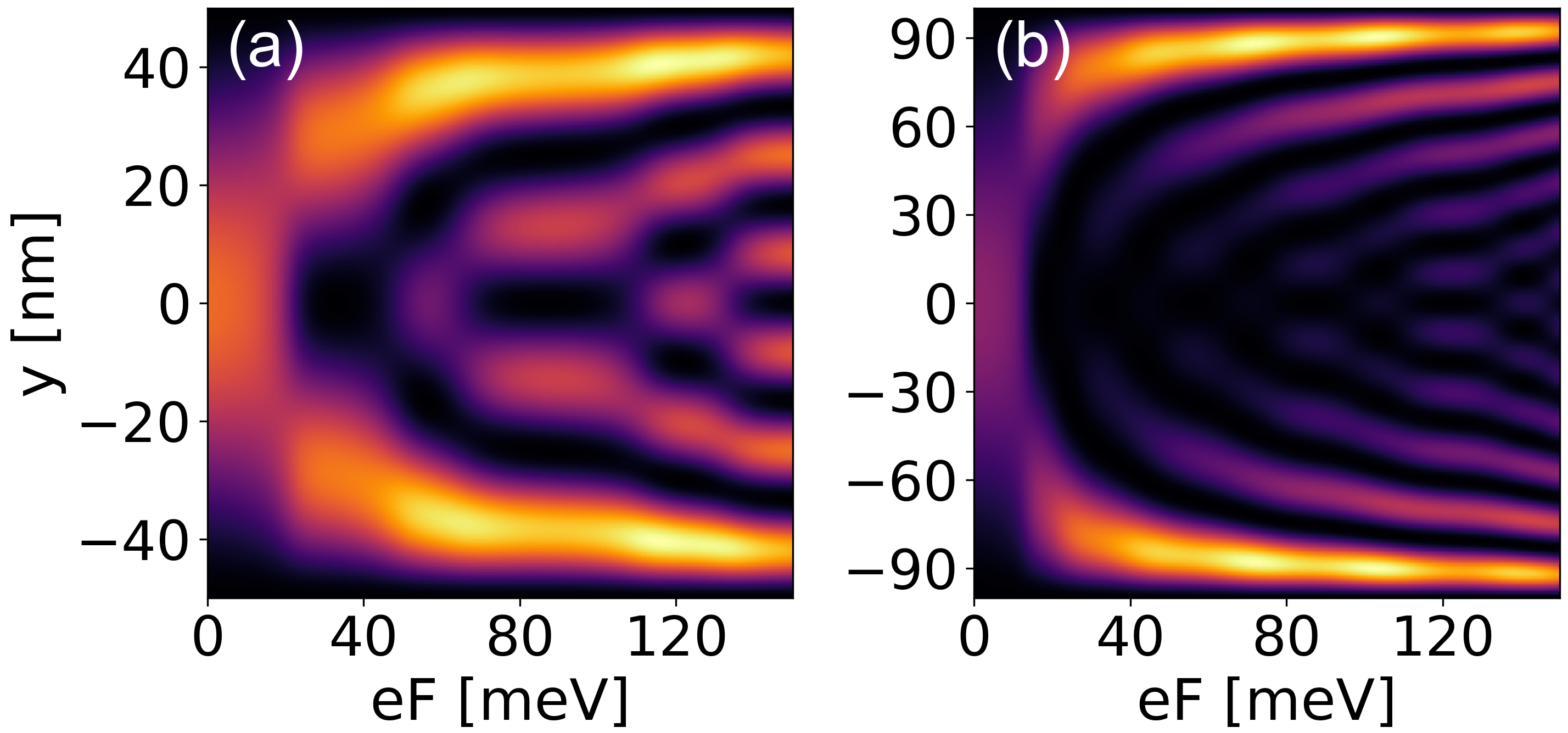}
        \caption{Electrically enhancement of the edge state localization for (a) $L_y = 100$ nm and (b) $L_y = 200$ nm.
        The color map indicates the transverse profile of the absolute squared value of the wave functions. In both panels, the wave functions adopted are those associated to the states represented by the blue line in the figure \ref{fig:band_structure_vsk}.}
        \label{fig:colormap_wavefunction}
    \end{center}
\end{figure}

Figure \ref{fig:colormap_wavefunction} shows the evolution of $|\psi(y)|^2$ as a function of the the electric field ($eF$) for $L_y = 100$ nm (Fig. \ref{fig:colormap_wavefunction}-a) and $200$ nm (Fig. \ref{fig:colormap_wavefunction}-b). In both cases, it becomes clear that the low-lying states tend to localize at the edges. In the narrow strip (Fig. \ref{fig:colormap_wavefunction}-a), the overlap between states at different edges leads to a stronger modulation as a function of the electric field.

The localization length $\xi(eF)$ can be extracted by fitting an exponential through the first two local maxima of $|\psi(y)|^2$ (a linear fit in a semi-log plot). Similarly, the wave number $k_f(eF)$ can be extracted by  by averaging the distances between subsequent minima of $|\psi(y)|^2$.

\begin{figure}[t]
    \begin{center}
        \includegraphics[width=1.0\columnwidth]{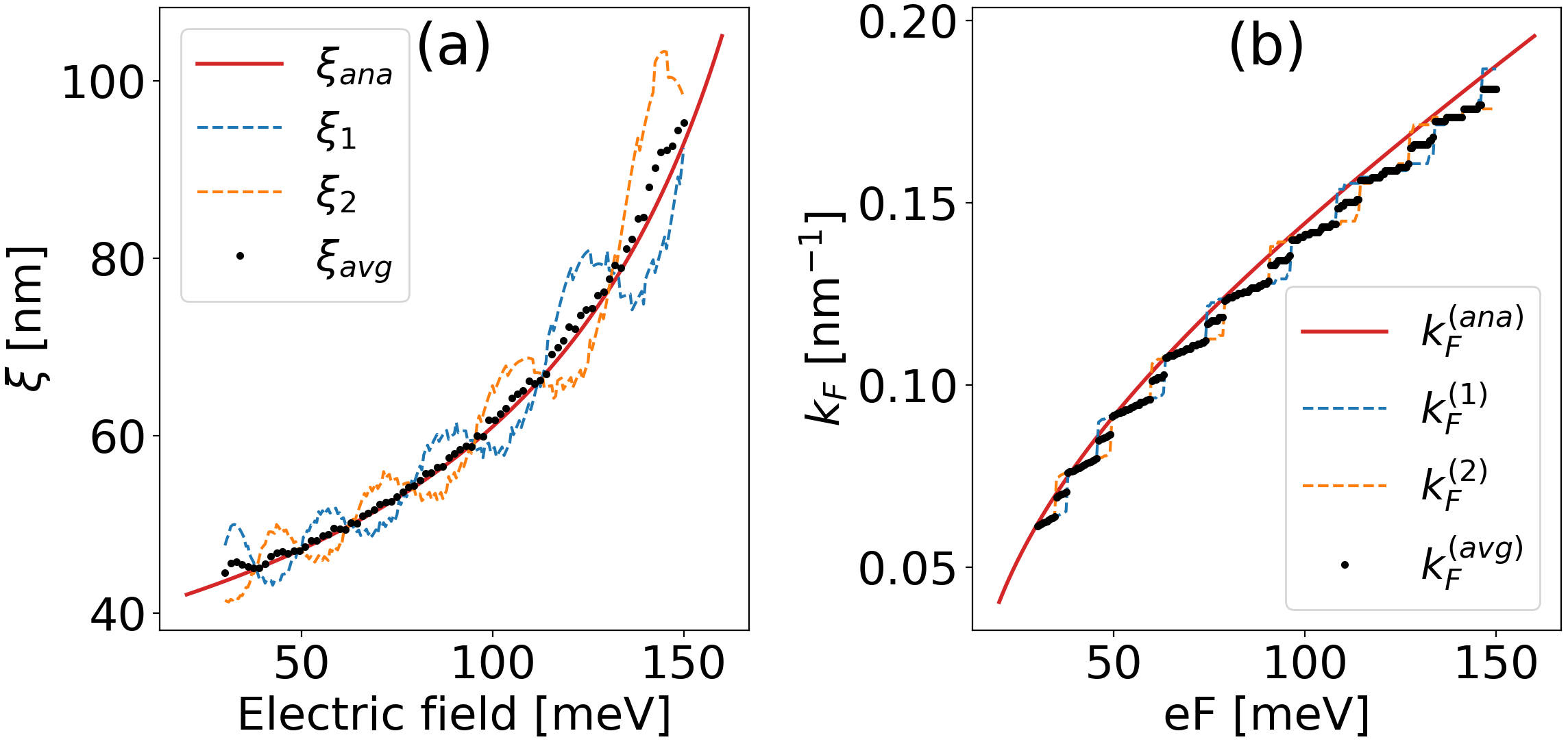}
        \caption{Evolution of the localization length $\xi$ (a) and the wave number $k_f$ (b) for the low-lying energy states for a system with $L_y = 200$ nm. In dashed lines we have the values obtained from numerical fitting over the wave functions, while the dotted values are the average of the results for both states and the continuum line is the analytical result.}
        \label{fig:xi_vs_eF}
    \end{center}
\end{figure}

In both cases, we can compare these fittings with analytical results (Eqs.~\eqref{eq:xi_eF} and \eqref{eq:kf_eF} in Appendix \ref{sec:appendix_analytical_solution}) obtained using the ansatz of by Eq.~\eqref{eq:absolute_squared_wave_function_model}. The results shown in Figure \ref{fig:xi_vs_eF} show an excellent agreement, further corroborating the choice of the ansatz.

\section{Concluding remarks}
\label{sec:Conclusions}

In summary, we studied the behavior of topological edge states in an realistic effective electronic model for GaSb/InAs quantum wells in the presence of an applied electric field. Using a ${\bf k} \cdot {\bf p}$ approach, were able to derive a realistic low-energy BHZ-like model for the system and probe the electric-field driven topological transition of quantum spin Hall phase of GaSb/InAs quantum-well systems.

One of our main results is establishing that the electric-field-driven energy oscillations in the edge states in narrow systems are a clear signature of the onset of the topological phase. Such oscillations are related to intra-edge coupling of edge states. More interestingly, we are able to provide an ansatz for the edge states as a function of the electric field with the same formal structure of those found in exponentially localized Majorana edge modes as a function of the magnetic field. Moreover, we were able to determine how the localization length and wave vector behave as a function of the electric field.

We believe our results can motivate experimental search to investigate such  oscillations in edge state conductance experiments, which could be an additional element in  confirming for the presence of topologically-protected helical edge modes in GaSb/InAs quantum-well devices.

\begin{acknowledgments}
    M.H.L.M. acknowledges
    financial support from CNPq and DAAD. R.L.R.C.T. acknowledges
    financial support from CNPq (Graduate scholarship 141556/2018-8) and FAPESP (Grant no.  2019/11550-8).
    G.M.S. acknowledges financial support from CNPq (grants No. 408916/2018-4 and 308806/2018-2) and CAPES (CsF-Grant No. 88881.068174/2014-01). L.G.G.V.D.S. acknowledges
    financial support by CNPq (Research Grants 308351/2017-7, 423137/2018-2, and 309789/2020-6) and FAPESP (grant No. 2016/18495-4). 
\end{acknowledgments}

\appendix

\section{Analytical solution for the edge wave-function and hybridization energy}
\label{sec:appendix_analytical_solution}

In order to derive an analytic expression for the wave-function and energy, we first define the energy of the $k_x\!=\!0$ edge states for a infinitely wide strip, i.e. $E_{\infty}(eF)=\lim_{L_y\to\infty} E_{\rm edge}(k_x\!=\!0)$.
This can be obtained as an order-3 polynomial approximation from numerical calculations. For a wide strip, the edge state wavefunction $\psi(y)$ will obey $H_{2x2}\psi(y)=E_{\infty}(eF)\psi(y)$, which can be written as
\begin{equation}\label{eq:hpsi_zero}
\begin{bmatrix}
E_c-\gamma_c \partial^2_y -E_{\infty} & i P \partial_y\\
 i P \partial_y &E_v-\gamma_v \partial^2_y -E_{\infty} 
\end{bmatrix} \psi(y)=0 \; .
\end{equation}

Since these  states are exponentially localized at the edges, we can expand the edge mode in the form $\psi(y)\sim e^{\pm z y}$ with $z=i k_f -1/\xi$ with $\xi$ being the localization length.  The solution of $z$ must satisfy the quartic equation:
\begin{equation}\label{eq:det_zero}
\det\begin{bmatrix}
E_c-\gamma_c z^2 -E_{\infty} & i P z\\
i P z &E_v-\gamma_v z^2 -E_{\infty} 
\end{bmatrix}=0 \; .
\end{equation}

For $\xi > 0$, there are only two solutions for $z$:
\begin{equation}\label{eq:sol_z}
z_{\pm}=-\sqrt{\frac{\Omega\pm \sqrt{\Lambda -\Omega^2}}{2\gamma_c\gamma_v}}
\end{equation}
where $\Omega=-P^2+(E_c-E_{\infty}) \gamma_v +(E_v-E_{\infty}) \gamma_c$ and $\Lambda=4\gamma_c \gamma_v (E_c-E_{\infty})(E_v-E_{\infty})$. Both $k_f=\mbox{Im}[z]$ and $1/\xi=\mbox{Re}[z]$ can be rewritten in terms of $\Omega$ and $\Lambda$:
\begin{equation}
    k_f=\frac{1}{2}\sqrt{\frac{\sqrt{\Lambda}+\Omega}{\gamma_c \gamma_v}} \; ,
    \label{eq:kf_eF}
\end{equation}
\begin{equation}
    \xi=-\frac{4 \gamma_c \gamma_v k_f}{\sqrt{\Lambda-\Omega^2}} \; .
    \label{eq:xi_eF}
\end{equation}

We can classify the two solutions as wave functions $\psi_t$ and $\psi_b$ localized at the top ($y\!=\!L_y$) and bottom ($y\!=\!0$) edges respectively. We can write, say, $\psi_b(y)$ as:
\begin{equation}
\psi_b(y) = u e^{z y}\begin{bmatrix}
A_0\\B_0
\end{bmatrix} + v e^{z^* y}\begin{bmatrix}
A_1\\B_1
\end{bmatrix} \; .
\end{equation}

By imposing time-reversal symmetry and the boundary condition $\psi_b(y=0)=0$, it is clear that $u=-v=i$ and $A(B)_0=A(B)_1$. Thus, a solution of Eq.~\eqref{eq:hpsi_zero} assuming Eq.~\eqref{eq:det_zero} is given by:
\begin{equation}
A_{0(1)} = \frac{(E_v -z^2 \gamma_v -E_{\infty})}{P z}, B_{0(1)} = I \; .
\end{equation}

Taking all of this into account, $\psi_b(y)$ is given by:
\begin{align}
\psi_b(y) \propto e^{-y/\xi}\sin(k_f y) \; ,
\end{align}
and  $\psi_t(y)$ (localized at the top edge) will be given by
\begin{align}
\psi_t(y) \propto e^{-(L_y-y)/\xi}\sin(k_f (L_y-y)) \; .
\end{align}

Finally, the energy difference associated with inter-edge coupling of the modes can be calculated by taking 
\begin{equation}
\Delta E =\frac{\langle\psi_t| H |\psi_b\rangle}{\langle\psi_{edge}|\psi_{edge}\rangle}
\end{equation}
where $\ket{\psi_{edge}}=a_0\ket{\psi_t} +a_1\ket{\psi_b}$ with the constant $|a_0|^2+|a_1|^2=1$ and $\langle\psi_{edge}|\psi_{edge}\rangle = \kappa$ is the normalization factor. In the limit of large system, $L_y \gg \xi$, the normalization can be approximated as
\begin{equation}
 \kappa\approx \frac{k_f^3 \xi^3}{4(k_f+k_f^3 \xi^2)}
\end{equation}
and the energy becomes
\begin{equation}
	\Delta E \approx \frac{k_f L}{2\xi \kappa}e^{-L/\xi}(\frac{|A_0|^2}{|A_0|^2+1} \gamma_c+\gamma_v)  \sin(k_f L) \; .
\end{equation}

For the parameters of Table~\ref{table:nonlin}, this means that $\Delta E$ oscillates with the electric field as 
\begin{equation}
\Delta E\approx \frac{2 k_f L}{\xi^2}e^{-L/\xi}(\gamma_c+\gamma_v)  \sin(k_f L) \; ,
\end{equation}
with $k_f$ given by Eq.~\eqref{eq:kf_eF}.

This derivation is analogous to that of \cite{DasSarma:Phys.Rev.B:220506:2012} for Majorana bound states in topological nanowires.


%

\end{document}